# Two and three photon excited luminescence of single gold nanoparticles: Switching between plasmon- and electron-hole-pair emission by ultrashort laser pulses

*Frank Wackenhut[‡,1], Xiao Wang[2], Antonio Virgilio Failla[3] and Alfred J. Meixner[‡,1]*

[1] Eberhard Karls University Tübingen, Institute of Physical and Theoretical Chemistry, 72076 Tübingen, Germany

[2] School of Physics and Electronic Science, Hunan University, Changsha, Hunan 410082, P. R. China

[3] University Medical Center Hamburg-Eppendorf, UKE Microscopy Imaging Facility, 20246 Hamburg, Germany

Abstract:

In this work we use femtosecond laser pulses of 800 nm wavelength to excite and characterize the multiphoton luminescence emission of single gold nanoparticles. For excitation with 100 fs laser pulses we observe a two and three photon emission dominated by radiative electron hole pair recombination, while the emission is caused by radiative plasmon decay for excitation with 500 fs pulses. For single gold nanorods with different aspect ratios, we study the interplay between the particle plasmon and electron hole pairs, which enables us to develop a quantitative model to fully describe the two and three photon luminescence emission of single gold nanoparticles.

Photoinduced luminescence from bulk gold has first been reported in the pioneering experiments of Mooradian in 1969 [1] and was attributed to the radiative recombination of electron hole pairs but was only of minor importance due to its very low quantum yield of about $10^{-10}$. Boyd et al. found that surface roughness could enhance the luminescence by several orders of magnitude with respect to smooth gold.[2] In recent years the interest in gold nanoparticles as photostable and non-blinking single optical markers has steadily increased due to their large scattering and absorption cross section.[3-4] It was pointed out that their linear optical properties, like elastic scattering [5-10] or one photon luminescence,[11-16] are dictated by the particle plasmon (PP), which is a coherent oscillation of the conduction band electrons.[4, 14, 17] However, the role of the PP [18-19] as well as the radiative recombination of electron hole pairs and the interplay between the two processes is still under debate.[20-23] Indeed, it appears not enough to consider only either the PP or electron hole pairs to fully describe the emission process of gold nanoparticles but also the interplay between these two processes needs to be taken into account.[11-12] This interplay will also influence the dynamics of non-linear processes such as two-photon (2PL) or three-photon (3PL) induced luminescence. Nevertheless 2PL of gold nanoparticles has been already extensively utilized for in vitro and in vivo bio-nanotechnology assays [24-27] and hence the complete understanding of the underlying processes is beneficial and necessary for further exploitation in *e.g.* material science, bio imaging, microscopy and spectroscopy.

Here we investigate experimentally the non-linear luminescence excitation of single gold nanospheres (GNS) and nanorods (GNR) with femtosecond laser pulses. We find that the dynamics leading to non-linear luminescence depends critically on the duration of the excitation pulses and to some extent also on the particle geometry. In addition to the well-known plasmon-based two photon luminescence following excitation with 500 fs pulses we find a novel excitation scheme with 100 fs pulses leading to two-photon and three-photon luminescence caused by the creation of electron-hole pairs via two-photon respectively three photon inter-band excitation and subsequent radiative electron-hole recombination from the Fermi-level to the initial levels. For these conditions we find that non-linear emission of the PP plays only a minor role.

All measurements were performed with a home built confocal microscope described in the references.[6, 11] For non-linear excitation we used amplified ultra short laser pulses of 800 nm with a minimal duration of 100 fs (Coherent Vitesse coupled into a Coherent ReGA amplifier). The gold nanoparticles were spin coated on glass coverslides and their concentration was adjusted to obtain a sample with single and optically isolated nanoparticles. However, melting of the nanoparticles under strong illumination is one of the major difficulties that has to be overcome to investigate their non-linear emission properties. In order to ensure that the nanoparticles were not altered or melted by repeated pulsed illumination we covered them with a 200nm $SiO_2$ layer, which has three advantages: (1) better thermal coupling to the environment, (2) conserve the shape in case that the particle is partially melted, (3) ensure a homogeneous dielectric environment. This procedure allows to strongly reduce melting of the nanoparicles. Linear and non-linear emission spectra of single 80 nm gold nanoshperes (GNS) are shown in Fig. 1a,b.

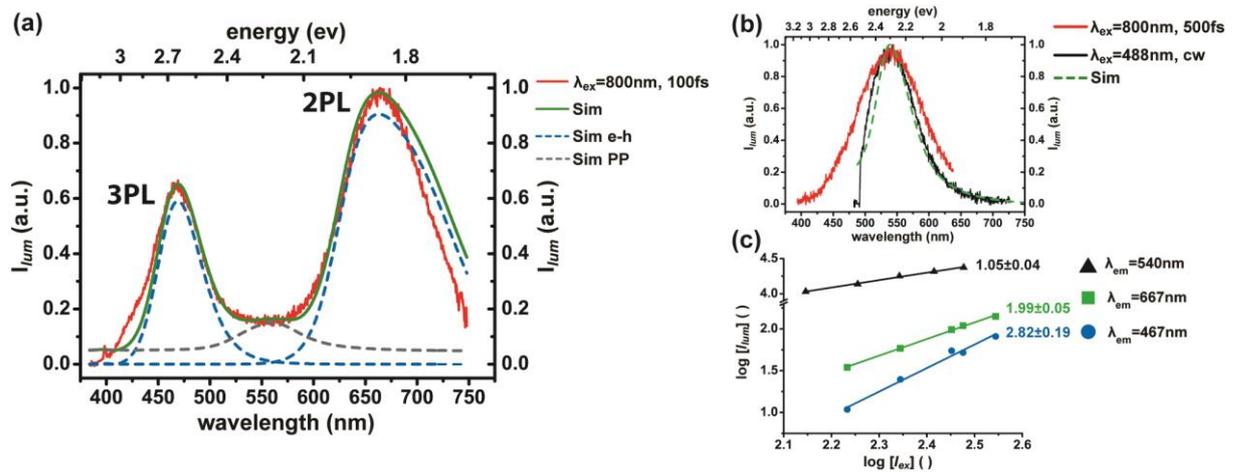

Figure. 1: (a) Non-linear luminescence spectrum (red line) of 80 nm GNS excited by 100 fs laser pulses at 800 nm showing two separated emission bands at $\lambda_{em}$=662 nm and $\lambda_{em}$=467 nm respectively. The exposure time was 30 seconds for all spectra shown in this work. The green line is a model spectrum with two separated bands for the luminescence from the electron hole recombination from the Fermi-level to the 5d- and 4d-levels, respectively (blue dashed lines). The dip between these two lines is partially filled due to radiative relaxation of the particle plasmon at 540 nm. In (b) the two photon luminescence induced (2PL) spectrum from the same particle excited at $\lambda_{ex}$=800 nm with a pulse duration of 500 fs is shown in red together with the (black line) one photon luminescence (1PL) spectrum excited with $\lambda_{ex}$=488 nm (continuous wave excitation) both having a spectral maximum at $\lambda_{em}$=540 nm. The maxima of both spectra overlap perfectly. In (c) the excitation power dependence of the respective spectral bands from (b) at $\lambda_{em}$=540 nm and from (a) at $\lambda_{em}$=662 nm and $\lambda_{em}$=467 nm are shown and the slopes of the linear fits indicate one, two and three-photon excited luminescence.

Figure 1a shows the non-linear luminescence spectrum (red line) of 80 nm GNS excited by 100 fs laser pulses at 800 nm consisting of two separated emission bands at $\lambda_{em}$=662 nm and $\lambda_{em}$=467 nm respectively. The repetition rate of the pulses was 250 kHz and an average power of 100 µW equivalent to a peak power in the focus of the objective (NA=1.25) of roughly 4.0 kW. As we will show below, this spectrum is caused by exciton recombination by an electron from the Fermi-level to the 5d and to the 4d levels respectively. In Figure 1b we show the non-linear luminescence spectrum when the duration of the laser pulses is increased from 100 fs to 500 fs duration reducing the pulse peak power to 0.8 kW and keeping the other excitation parameters constant. This spectrum has only one maximum located at $\lambda_{em}$=540 nm. As we can see the later spectrum is very similar to the one photon induced luminescence (1PL) spectrum (black line) of the 80 nm GNS excited with a continuous wave (cw) laser at $\lambda_{ex}$=488 nm, which is caused by radiative plasmon decay.[11, 19, 28] The green dashed line represents the simulated one-photon luminescence. The main difference between the two experimental spectra in Fig. 1b is the width of the non-linear spectrum which increases from 91 nm (0.39 eV, $\lambda_{ex}$=488 nm) found for the linear spectrum (black) to 125 nm (0.54 eV, $\lambda_{ex}$=800 nm), this translates into a reduction of the PP dephasing time from 3.4 fs to 2.4 fs.[5, 11, 29] This spectral broadening is caused by a faster dephasing (loss of coherence) of the PP due to the high temperature of the electrons close to the Fermi level after the excitation with an intense laser pulse.[9, 29-31]

Interestingly, the non-linear luminescence profile obtained with the 100 fs pulses (Fig. 1a) looks different to the one obtained with the 500 fs pulses (Fig. 1b) and has two distinct spectral emission maxima, located at $\lambda_{em}$=662 nm and $\lambda_{em}$=467 nm. The red shifted emission maximum was already assigned to the two photon luminescence (2PL),[20-21, 23] but the emission at $\lambda_{em}$=467 nm has, to our knowledge, not been reported yet for single gold nanoparticles. A plasmonic origin of these two emission peaks, like for the spectra in Fig. 1b, can be excluded because the PP resonance of a GNS is spectrally located between these two emission peaks. The increase of the temperature of the conduction band electrons close to the Fermi level is larger after the excitation with a shorter laser pulse (compared to the pulse duration used in Fig. 1b) and leads to an even faster dephasing of the PP. This effect can be seen as a reversible bleaching of the PP and has already been observed in transient absorption experiments.[32-34] Hence radiative PP decay is not sufficient to explain this emission band. To gain further insight, we have studied the excitation power dependence of the spectral signals displayed in Fig. 1a and the results are shown in Fig. 1c in a double logarithmic representation, where the slope of the linear fits depicts the order of the underlying process. The black data points in Fig. 1c are obtained from the 1PL spectrum shown in Fig. 1b excited with a cw laser at $\lambda_{ex}$=488 nm and the slope is close to unity confirming the expected linear power dependence. Data points shown in blue and green are obtained from the spectra excited with 100 fs pulses where the emission peaks are spectrally located at $\lambda_{em}$=467 nm (blue data points) and $\lambda_{em}$=662 nm (green data points) in Fig. 1a. The slope of the linear fit for the emission at $\lambda_{em}$=662 nm is 1.99±0.05, which shows that this emission is due to 2PL, as was already described in literature.[20-21, 23] Importantly for the emission peak at $\lambda_{em}$=467 nm we find a slope of 2.82±0.92 in the linear fit in Fig. 1c showing that the respective excitation is a three photon process (3PL), which only can be observed with a short pulse duration indicating that the underlying process is a simultaneous absorption of three photons.

In the following we present a quantitative model based on two photon interband excitation from the 5d to the 6sp band and three photon interband excitation from the 4d the 6sp band, respectively, and subsequent radiative recombination of electrons from the Fermi level with the respective holes in the d-bands which is suitable to explain the main features of the non-linear luminescence emission observed in Fig. 1a.

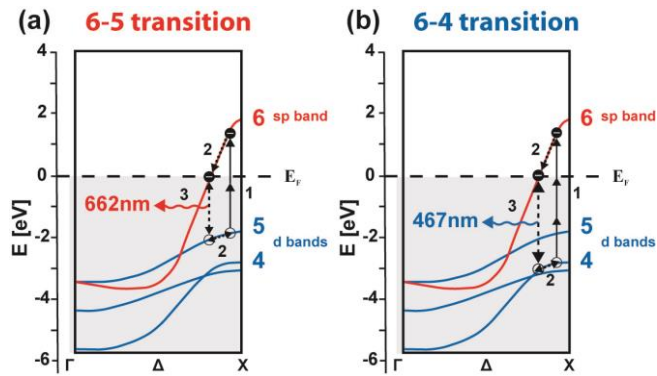

fig. 2: Energy level diagram to explain the two photon (a) and three photon (b) excitation and subsequent emission: Excitation of a d-band electron by two (a) or three (b) photons near the X symmetry point creates an electron hole pair. The second step is the relaxation of the excited electron hole pair followed by radiative recombination. For more details on the relativistic band structure of gold close to the X symmetry point see ref. [31].

The band structure of gold close to the X symmetry point shown in Fig. 2 is adopted from Eckardt et al..[35] The energy gap is, according to the calculations reported in Boyd et al.,[2] 1.86 eV (662 nm) and 2.59 eV (467 nm) between the Fermi level and the d-bands (labelled as 5 and 4 in Fig. 2). These energy gaps are in perfect agreement with the spectral positions of the 2PL and 3PL peaks observed in Fig. 1a and can be used to explain both the 2PL and 3PL caused by the radiative recombination of electron hole pairs with a simple three step model, which is similar to the one presented in reference[2] where it is used to describe 1PL of gold surfaces. In details, step 1: excitation of an electron above the Fermi level $E_f$, which creates an excited electron hole pair. Step 2 is the relaxation of this electron hole pair, which subsequently recombines radiatively (step 3). Note that the d-band holes (step 1) can be created in the d-bands labelled as 5 or 4 by the absorption of two or three photons resulting in either 2PL or 3PL. These three steps can in principle occur at different positions in the Brillouin zone (*e.g.* close to the L symmetry point), but only the radiative recombination close to the X symmetry point results in the emission wavelengths observed in Fig. 1a. Energy transfer to a PP or other non-radiative processes are not displayed in this scheme, but need to be considered to yield a quantitative description of the involved processes.

The corresponding calculation can be separated into two parts. Firstly, we calculate the luminescence emission caused by radiative recombination of electron hole pairs following Boyd et al.[2] which can be described as:

$$I_{e-h} \propto F(\omega_1, \omega_2) \int dE \, [D(E, \hbar\omega_2) \, f_e(E) \, f_h(E_h)],$$

where $\omega_1/\omega_2$ are the excitation/emission frequencies, $D(E, \hbar\omega_2)$ is the density of states based on the relativistic band structure of gold close to the X-symmetry point of the first Brillouin zone and $f_e(E)/f_h(E_h)$ are the Fermi Dirac distributions of electrons/holes. More details are given in the supporting information. The calculated spectral peak positions for a 6-5 and a 6-4 transition (blue dashed lines in Fig. 1a) agree perfectly with the experimental data in peak position and bandwidth, while the intensity was adjusted for representation purposes to fit the experimental data. The intensity in between these two luminescence bands can naturally be explained by the radiative decay of the PP according to reference:[36]

$$I_{PP} \propto \frac{\varepsilon''(\omega_1) \, k_1}{C_{abs}(\omega_1)\varepsilon_0} \int \frac{dV \, R(\omega_1,r)R(\omega_2,r)}{q_d[P(\omega_2,r)+R(\omega_2,r)-1]+1},$$

where $R(\omega_2,r)/P(\omega_2,r)$ are the radiative/non radiative decay rate, $C_{abs}(\omega_1)$ is the absorption cross section, $\varepsilon''(\omega_1)$ is the imaginary part of the dielectric constant of gold, $k_1$ the wave vector of the excitation light and $q_d$ is the quantum efficiency of bulk gold. The result of this calculation is plotted as dashed grey curve in Fig. 1a. With the amplitudes of the three contributions as only free parameter we can perfectly model measured spectrum as can be seen by the green line in Fig. 1a. We note that the by far dominant contribution to the experimental spectrum in Fig. 1a is caused by the radiative recombination of electron hole pairs excited by a two-photon and three-photon absorption process. Furthermore, the contribution of the PP is comparably weak, which can be understood as its dephasing time is strongly decreased after the excitation with an intense laser pulse.

In order to further prove that these emission peaks are related to the band structure of gold (radiative recombination of electron hole pairs) and only weakly influenced by the PP we acquired the non-linear luminescence spectra of single gold nanorods (GNRs), which are shown in Fig. 3.

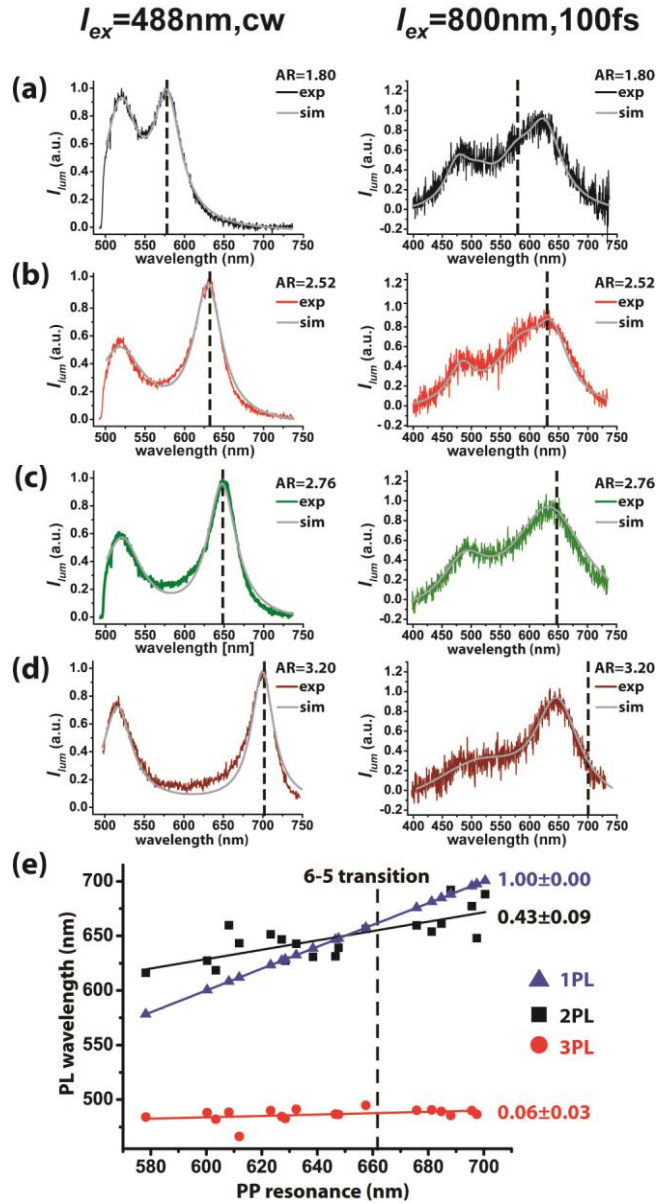

Fig. 3: One-photon luminescence spectra of single GNRs are shown in the left column and illustrate the different PP resonances (indicated by the dashed lines) due to different aspect ratios; $\lambda_{em}$=577 nm (a), $\lambda_{em}$=630 nm (b), $\lambda_{em}$=651 nm (c) and $\lambda_{em}$=691 nm (d). The right column shows the corresponding non-linear emission spectra of the same GNRs with the 2PL and 3PL contribution. The grey lines do show corresponding fits. (e) Spectral position of luminescence emission plotted against PP resonance showing only a weak influence of the PP.

GNRs are chosen for this experiment since the plasmon resonance depends on their aspect ratio (ratio long/short axis), the spectral position of the longitudinal PP resonance shifts to longer wavelengths with increasing aspect ratio whereas the transverse PP resonance stays at short wavelengths and the emission maximum shifts only by a small amount.[11] This effect can be observed in the first column of Fig. 3, where the 1PL spectra excited at $\lambda_{ex}$=488 nm are shown for single GNRs with a fixed short axis length of 25nm (specified by the manufacturer, Nanopartz) and a variable length of the long axis. The respective longitudinal PP resonances (indicated by the dashed lines in Fig. 3 are $\lambda_{em}$=577 nm (a), $\lambda_{em}$=630 nm (b), $\lambda_{em}$=651 nm (c) and $\lambda_{em}$=691 nm (d) giving aspect ratios ranging from 1.8 in (a) to 3.2 in (d). The same GNRs are excited at $\lambda_{ex}$=800 nm with a pulse duration of 100 fs and their corresponding non-linear luminescence spectra are presented in the second column of Fig. 3. Indeed, the spectra of the GNRs excited at $\lambda_{ex}$=800 nm exhibit 2PL and 3PL emission peaks at approximately $\lambda_{em}$=650 nm and $\lambda_{em}$=470 nm regardless of their aspect ratio. These values are quite similar to the ones already observed for GNSs in Fig. 1a, which indicates that the observed luminescence emission only slightly depends on the particle geometry (aspect ratio). Furthermore, the 2PL and 3PL emission maxima are not necessarily located at the PP resonance of the respective GNR which is illustrated in Fig. 3e. Here the spectral position of the luminescence emission maxima is plotted against the spectral position of the PP. The blue, black and red data points present 1PL, 2PL and 3PL, respectively. In this representation a slope of zero corresponds to the case when the PP has no influence on the luminescence emission and hence the emission peak is located at the same spectral position for all particles, which is given by the band structure of gold. In contrast the blue line (emission maxima of the 1PL) presents the case when the emission is located at the PP resonance, which will give a line through the origin with a slope of one. Therefore, the slope in this representation is a measure for the influence of the PP on the luminescence emission, as a slope of zero shows that there is no correlation between the PP and the luminescence emission, while a slope of one represents the situation where the emission is fully determined by the PP resonance. The first case is observed for the 3PL, presented by the red line in Fig. 3e, where the slope is close to zero (0.06±0.03). This demonstrates that the 3PL is always located at the same spectral position given by the band structure and is independent of the particle AR and therefore is not influenced by the PP. The black line in Fig. 3e is obtained by analyzing the 2PL and has a slope larger than zero (0.43±0.09). However, the black (2PL) and blue (1PL) line do not coincide and the slope is between zero (no influence of the PP) and one (fully determined by the PP), which shows that the PP has an influence on the 2PL but the emission is not fully dictated by it. In Fig. 1a it has been shown that the emission caused by the 6-5 transition is located at approximately 660 nm and hence is in the same spectral region like the PP resonance for some of the GNRs used in this experiment. This spectral overlap causes an interaction between the PP and the electron hole pair and leads to the observed influence of the PP on the 2PL. A similar effect was already observed for the 1PL of GNRs.[11] However the impact of the spectral overlap between the electron hole pair transition and the PP on the luminescence emission is also evident for the 3PL, as there exists no spectral overlap between the PP and the 6-4 transition and no influence of the PP on the luminescence emission can be observed. This emission behavior underlines that the non-linear emission properties of gold nanoparticles can only be understood if one considers the interplay between the PP and electron hole pairs.

In conclusion we have shown that the emission properties of a single 80 nm diameter GNS can either be based on the radiative decay of PP for long pulse durations (500 fs) or on the radiative recombination of electron hole pairs for short pulse duration (100 fs). This shows that the luminescence emission of gold nanoparticles is a highly dynamic process and explains why some researchers observed an emission based on the PP,[12,13] while others associated its origin with electron hole pairs.[20-23] Furthermore we report for the first time a strong luminescence excited by three photons triggered by a transition from the 4d-band to the 6sp-band of a single gold nanoparticle, which can be detected for both single GNSs and GNRs. In addition, we show that the two and three photon induce luminescence only weakly depends on the particle geometry and that the spectral position of the emission maximum is quite similar for GNSs and GNRs with different aspect ratios. This indicates that the main component of the non-linear emission originates indeed from the band structure and is caused by the radiative recombination of electron hole pairs. This observation is also supported by a quantitative model explaining the two and three photon luminescence excitation with a simple three step process, which involves the excitation, relaxation and radiative recombination of electron hole pairs. Finally, this model considering the interplay between the PP and electron hole pairs can perfectly explain the multiphoton emission spectra of gold nanoparticles.